\documentclass[preprint,preprintnumbers,amsmath,amssymb]{revtex4}
\usepackage{amsfonts}    
\usepackage{amssymb}
\usepackage{latexsym}
\usepackage{graphicx}
\usepackage[usenames,dvipsnames]{color}
\usepackage{tikz}
\usetikzlibrary{3d,calc}

\newcommand{\be}{\begin{equation}}
\newcommand{\ee}{\end{equation}}

\newcommand{\pa}{\partial}

\newcommand{\la}{\lambda}

\newcommand{\om}{\omega}

\newcommand{\rar}{\rightarrow}

\newcommand{\non}{\nonumber}
\begin{document}

\title{Trans-series for the ground state density and Generalized Bloch equation}

%

\author{E.~Shuryak$^1$}
\email{edward.shuryak@stonybrook.edu}

\author{A.V.~Turbiner$^{1,2}$}
\email{turbiner@nucleares.unam.mx, alexander.turbiner@stonybrook.edu}


\affiliation{$^1$  Department of Physics and Astronomy, Stony Brook University,
Stony Brook, NY 11794-3800, USA}

\affiliation{$^{2}$ Instituto de Ciencias Nucleares, Universidad Nacional Aut\'onoma de M\'exico,
Apartado Postal 70-543, 04510 M\'exico, D.F., M\'exico}

\begin{abstract}
Based on Generalized Bloch equation the trans-series expansion
for the phase (exponent) of the ground state density for double-well potential is constructed.
It is shown that the leading and next-to-leading semiclassical terms are still defined by the flucton trajectory (its classical action) and quadratic fluctuations (the determinant), respectively, while the
the next-to-next-to-leading correction (at large distances) is of non-perturbative nature. It comes from the fact that all flucton plus multi-instanton, instanton-anti-instanton classical trajectories lead to the same classical action behavior at large distances! This correction is proportional
to sum of all leading instanton contributions to energy gap.
\end{abstract}

\maketitle

It has been understood long ago that the inter-relations between two formulations of quantum mechanics,
Schr\"odinger's based on the wave functions and Feynman's based on path integrals, becomes non-trivial
in certain special problems. In particular, if coordinates are defined on compact manifolds
(such as Lie groups), there exists topologically distinct paths. Since they cannot be
continuously deformed into basic topologically trivial paths, the issue of their normalization
(and especially their sign) in the path integral formalism is non-trivial and requires basically
a separate definition. It has been very clearly explained in the remarkable paper by L.~Schulman
\cite{Schulman} using the simplest example of a particle on a circle (or $O(2)=U(1)$ group),
in which case the question is whether angular momentum should be integer or half-integer.
In the latter case the wave functions must be defined as anti-periodic, and the winding paths
contribution to the integral as having an extra sign factor. Only with it, the path integral
formalism had become finally fixed uniquely.

In our previous works \cite{Escobar-Ruiz:2016,Escobar-Ruiz:2017} we introduced and studied a version
of the semiclassical theory based on the so called $flucton$ paths in Euclidian time,
the periodic ones which start and end at some arbitrary location $x_0$ and thus contributing
to the density matrix $\rho(x_0)$. Unlike the textbook WKB approach, this one can
be used for multidimensional or QFT problems, and
perturbative corrections to all orders can be calculated  via Feynman diagrams.
These corrections  has been explicitly calculated, in one and  two loops for a number of
examples including quartic anharmonic oscillator and sine-Gordon potential.
These series on top of flucton were then reinterpreted and rederived, using
the so-called generalized Bloch equation.

If the potential of the problem has a single minimum, like in anharmonic oscillator $V \sim x^4$
the flucton path is uniquely defined by a condition that at the Euclidian time
$\tau\rightarrow \pm \infty$ it
should ``relax" to that minimum. However, if there are two or more degenerate minima (as is the case
in the double-well or sin-Gordon problems we also studied), there are also paths which can  ``relax"
to two different minima. Classical paths, corresponding to transitions between those minima are known
as $instantons$ (or anti-instantons, or multi-instantons in general). Contributions of instantons
to the ground state energy has been studied in multiple papers, including e.g. our own works
\cite{EST-I,EST-II} where it also has been done explicitly, up to 3 loops.

\begin{figure}[htbp]
\begin{center}
\includegraphics[width=3.5cm]{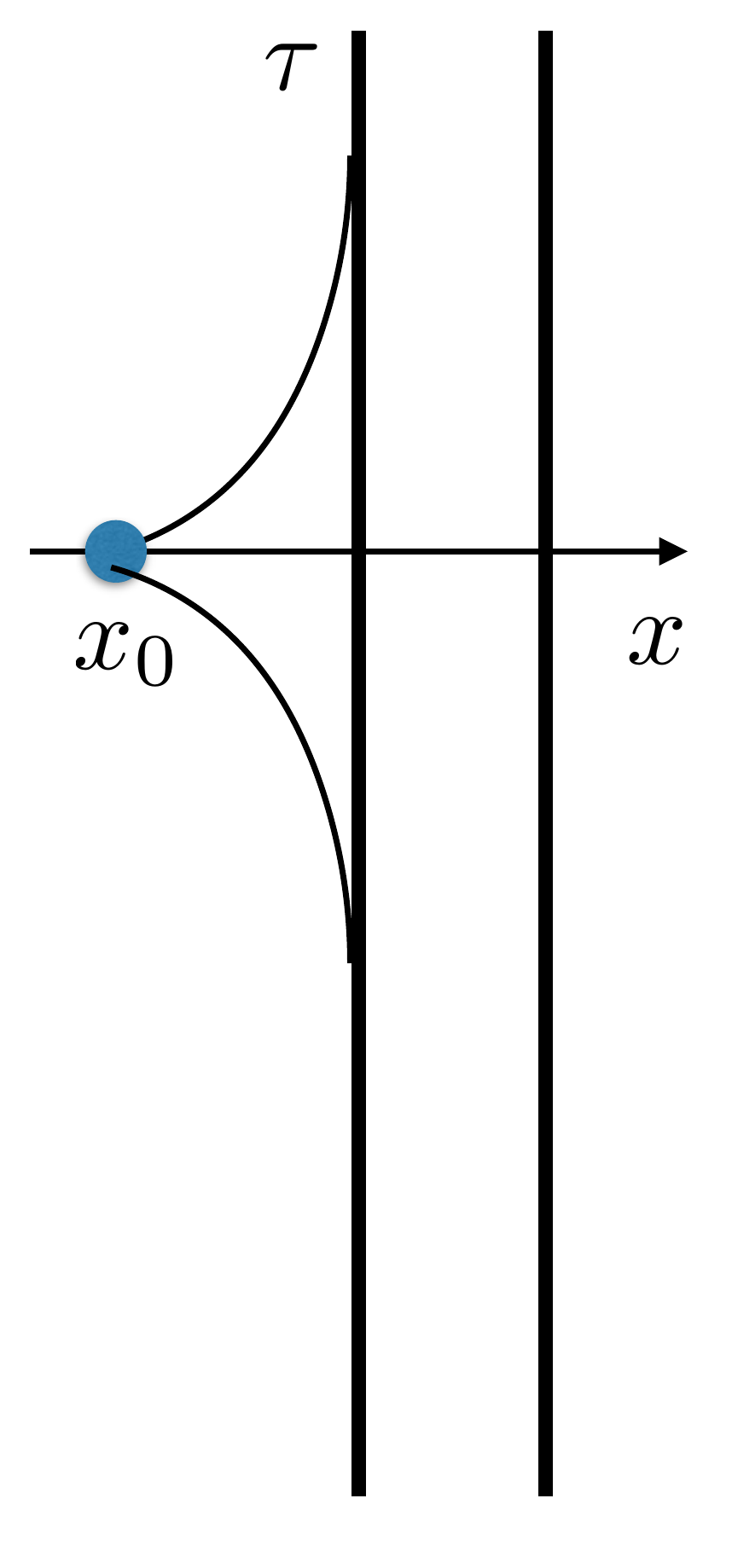}
\includegraphics[width=3.5cm]{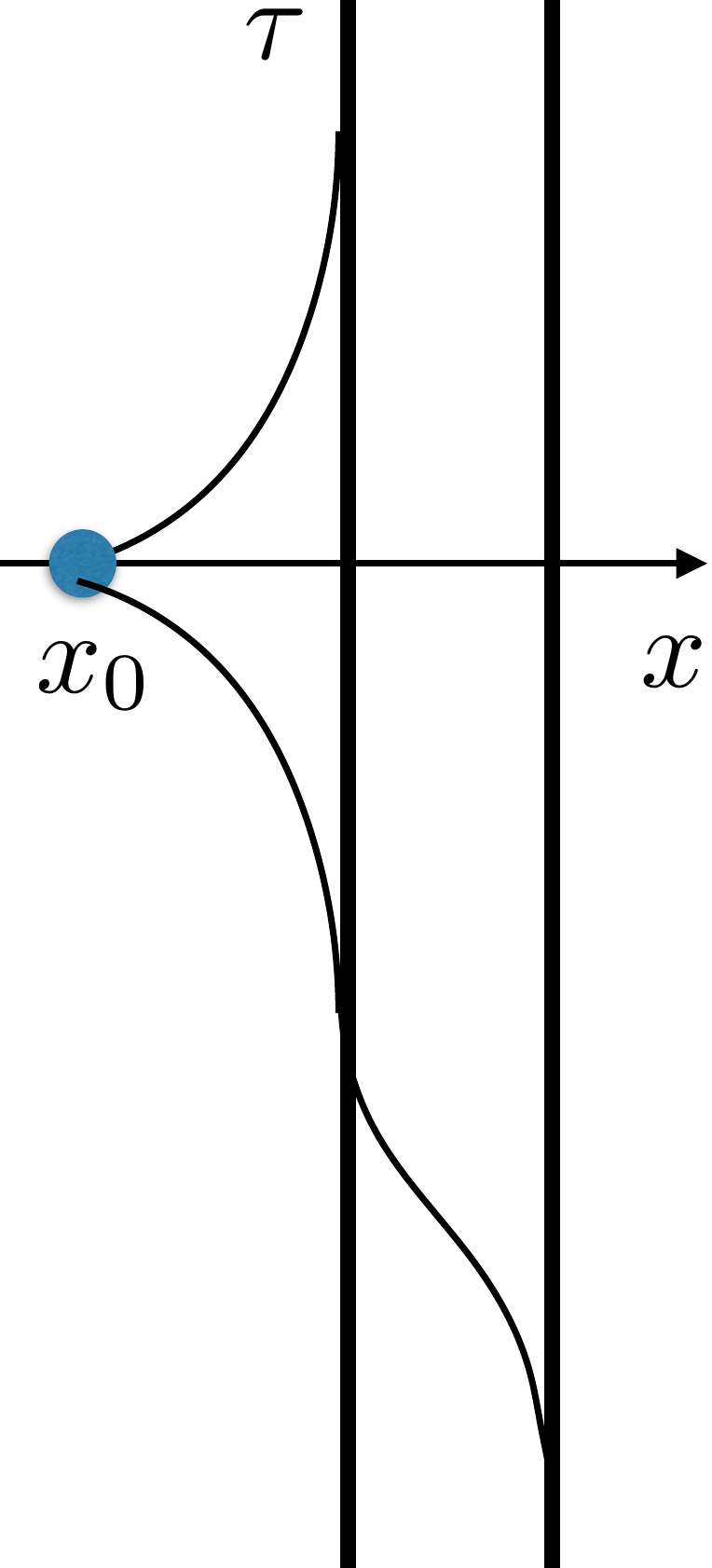}
\caption{The flucton path (left) and flucton-plus-instanton path (right)
both pass form some generic point $x_0$ and relax to one or two degenerate minima,
to ensure the finiteness of the action. }
\label{fig_fluct_inst}
\end{center}
\end{figure}

The issue we address in this work is the instanton contribution to
the the density matrix. In Fig. \ref{fig_fluct_inst} we illustrate it by two paths,
both passing through some generic point $x_0$ (which we take to be outside of both
potential minima marked by wide solid lines). The left sketch shows the flucton path,
which at $\tau\rightarrow \pm \infty$ relaxes to the same (nearest) minimum.
The right sketch shows a path  which  relaxes to different minima: we will call it ``f+i"
{\em (flucton plus instanton)} path. The Euclidean time $\tau$ is the vertical coordinate.
(Recall that at finite temperatures it is defined on a circle with circumference $\beta=\hbar/T$,
and the paths should be periodic. Yet in this work we consider zero temperature quantum mechanics,
so $\beta=\infty$ and the only remaining condition is that the paths must have a finite action.)

Since both paths pass through the point $x_0$ they both must contribute to  $\rho(x_0)$.
Yet since the paths are topologically distinct, the question of  relative normalization of their
contributions to the integral naturally arises.
We already touched upon this issue in our previous paper \cite{EST-II} (for $0 < x_0 < 1$ in between
the minima) but now we would like to do it more explicitly, using the classic example of the double
well potential and the generalized Bloch equation we also introduced before \cite{EST-II}.

Nowadays it is well known fact that in quantum mechanics for potentials with two or more degenerate minima the ground state energy contains non-analytic terms at $g \rar 0$ of instanton origin in addition to perturbation theory in $g$, see for instance \cite{Polyakov:1977}. In particular, for the ground state of the celebrated quartic double-well potential the standard perturbation theory expansion for energy becomes trans-series of the form,
\begin{eqnarray}
E(g^2)& = &  E_{\rm PT}(g^2)
\nonumber\\[3mm]
      & + &  \sum_{k=1}^\infty \sum_{l}\sum_{p=0}^\infty
\underbrace{ \left(\frac{1}{|g|}\,\exp\left[-\frac{c} {g^2}\right]\right)^k }_{ \rm k-instanton}\,
{ \left(\log \frac{c}{g^2}\right)^l}\, \underbrace{c_{k, l, p} \  g^{2p}}_{\rm  PT} \ ,
\label{trans}
\end{eqnarray}
see e.g. \cite{Shifman:2015},
where the parameter $c=1/6$ and $c_{k, l, p}$ are real parameters, and $g$ is the coupling constant (see below), the subscript PT stands for perturbation theory. Similar expansion can be derived for all energy eigenvalues. Perhaps, L.D.~Landau and E.M.~Lifschitz were the first who indicated to this phenomenon \cite{LL}, J.~Zinn-Justin \cite{Zinn-Justin:1981} derived this expansion systematically as a state-of-the-art and together with U.~Jentschura \cite{ZJJ:2004} they made impressive concrete calculations of this expansion. Recently, Dunne-\"Unsal in a number of papers revealed the hidden properties of (\ref{trans}) and made it understandable, at least, for present authors, see e.g. \cite{Dunne-Unsal:2014} and references therein. Note that (\ref{trans}) implies that the energy can be written as sum of perturbative and non-perturbative parts,
\begin{equation}
\label{EPT+ENPT}
  E\ =\ E_{\rm PT}\ +\ E_{\rm NPT}\ .
\end{equation}

The aim of this paper is to derive non-analytic terms in $g$ for ground state density (the square of the ground state function) in a systematic way, thus, constructing a type of trans-series for wavefunction assuming that the trans-series for the ground state energy is known.
Explicitly, it is done by separating perturbative and non-perturbative parts in wavefunction {\it multiplicatively},
\begin{equation}
\label{psiPT+psiNPT}
  \Psi\ =\ e^{-\phi_{\rm PT}\ -\ \phi_{\rm NPT}}\ \equiv \ \psi_{\rm PT}\,\psi_{\rm NPT}\ ,
\end{equation}
hence, the log of wavefunction can be represented as sum of perturbative and non-perturbative terms. This is the key observation which comes naturally from the Riccati-Bloch equation.
Then we will try to clarify the obtained trans-series in the framework of path integral formalism. The celebrated quartic double-well potential will be taken as the example.
Thus, overall, the derivation will be made from two different directions: (i) from quantum mechanics using the the generalized Bloch equation of the type presented in \cite{Escobar-Ruiz:2017} and (ii)
from the Euclidian time path integral following a variety of flucton-instanton trajectories.

Needless to say that the celebrated quartic double-well potential, written for the future convenience in the form
\begin{equation}
\label{DWP}
                 V(x)\ =\ \frac{1}{2}\,x^2 (1 - g x)^2\ ,
\end{equation}
where $g$ is the coupling constant, plays exceptionally important role in different physical sciences and chemistry. It has two degenerate minima situated at $x=0$ and $x=\frac{1}{g}$, respectively, and maximum at $x=\frac{1}{2g}$. The potential is also symmetric with center of symmetry at $x_c=\frac{1}{2g}$,
\[
V(x - \frac{1}{2g})\ =\ V(- x + \frac{1}{2g})\ ,
\]
It is seen explicitly when the potential (\ref{DWP}) is rewritten as
\begin{equation}
\label{DWP-alt}
                 V(\tilde x)\ =\ \frac{g^2}{2}\,(\tilde x-\frac{1}{2g})^2
                 (\tilde x +\frac{1}{2g})^2\ -\ \frac{1}{32 g^2}\ ,
\end{equation}
where $\tilde x=x - \frac{1}{2g}$.
It implies the parity of the eigenfunction, being even or odd. Hence, eigenfunction can be represented in the form
\begin{equation}
\label{e-DWP}
     \Psi(x)\ =\   \Psi(x - \frac{1}{2g})\ \pm \  \Psi(- x + \frac{1}{2g})\ ,
\end{equation}
with sign plus for even and sign minus for odd eigenfunctions 
\footnote{In folklore it is known as the E.M.~Lifschitz prescription. Taking for instance $\Psi(x)=e^{-g^2 x^2}$ the energy gap can be evaluated up to the multiplicative constant}.
However, to study the trans-series expansion in quantum mechanics for the ground state eigenfunction it is more convenient to use the exponential representation (\ref{psiPT+psiNPT}) where the phase is given by sum of perturbative and non-perturbation parts,
\[
  \log \Psi\ =\ \log \Psi_{PT} + \log \Psi_{NPT}\ .
\]
It is evident that the in QFT the path integral for the density matrix in saddle-point method the representation (\ref{e-DWP}) is more natural, it appears as the sum of saddle-point contributions for large positive (negative) distance $x_0$.

The potential (\ref{DWP}) belongs to a special class of anharmonic potentials
\be
\label{potential}
   V(x)\ =\ \frac{\tilde V(g x)}{g^2}\ =\ \frac{1}{2}\, x^2 + a_3\, g x^3 + a_4\, g^2 x^4 + \ldots\ ,
\ee
as well as celebrated sine-Gordon potential, where $\tilde V$ has a minimum at $x=0$; it always starts from quadratic term, the frequency of the small oscillations near minimum can always be placed equal to one, $\om=1$ and $g$ is the coupling constant of dimension $[\frac{1}{x}]$, see e.g. \cite{{Escobar-Ruiz:2017}}. For the sake of future convenience, the classical (vacuum) energy is always taken to be zero, $V(0)=0$, and $a_{2,3,\ldots}$ are real, dimensionless parameters, hence, $V(x) \geq 0$. We call $(g x)$ the {\it classical} coordinate, see below. Both the classical coordinate and the Hamiltonian with the potential (\ref{potential}),
\be
\label{Hamiltonian}
{\cal H}\ =\ -\frac{1}{2m} \pa_x^2 \ +\ \frac{1}{g^2}\,\tilde V(gx)\ ,\ \pa_x=\frac{d}{dx} \ ,\ m=1\ ,\ x \in (-\infty, \infty)\ ,
\ee
are invariant with respect to simultaneous change
\[
     x \rar -x \quad , \quad g \rar -g\ .
\]
It implies that the energy is the function of $g^2$,
\be
\label{energy}
    E\ =\ E(g^2)\ .
\ee

A particular form of the trans-series (\ref{trans}) for the ground state energy of the quartic double-well potential (\ref{DWP}), which we are going to exploit, has the form (if for a sake of simplicity we assume $g>0$),
\be
\label{E-PT-Trans}
     E(g)\ = \ E_{\rm PT}\ +\ E_{\rm NPT}\ =\ \sum_{n=0} g^{2n} E_{PT,n}\ +
\ee

\[
 \frac{1}{g}\,e^{-S_0} \, ( A^{(1)}_0 + A^{(1)}_1 g^2 + \ldots)\ +\
 \frac{1}{g^2}\,e^{-2S_0}\,( A^{(2)}_0 + A^{(2)}_1 g^2 + \ldots) + \ldots
\]

\[
 + \log (g^2)\,\frac{1}{g}\,e^{-S_0} \bigg(\frac{1}{g}\,e^{-S_0}\,(B^{(1)}_0 + B^{(1)}_{0,1} g^2 + \ldots) +
  \frac{1}{g^2}\, e^{-2S_0} (B^{(1)}_1 + B^{(1)}_{1,1} g^2 + \ldots) + \ldots \bigg)\ +
\]

\[
 + \log^2 (g^2)\frac{1}{g^2}\,e^{-2S_0} \, \bigg(\frac{1}{g}\,e^{-S_0}(B^{(2)}_0 + B^{(2)}_{0,1} g^2 + \ldots) + \frac{1}{g^2}\,e^{-2S_0} (B^{(2)}_1 + B^{(2)}_{1,1} g^2 + \ldots)  + \ldots \bigg)\ + \ldots\ ,
\]
where $S_0=\frac{1}{6 g^2}$ is one-instanton classical action,
the parameters $E$'s, $A$'s and $B$'s are real and can be calculated constructively,
and some of them are explicitly known, see \cite{ZJJ:2004} and references therein. The form (\ref{E-PT-Trans}) is slightly different from the standard form of trans-series,
see e.g. \cite{Dunne-Unsal:2014}, being of the type  (\ref{trans}): it takes into account
the appearance in the standard form for trans-series the imaginary parts in some
coefficients with their further cancellations due to Bogomolny mechanism \cite{Bogomolny:1980}
\footnote{In standard calculations of the exponentially small terms the logarithmic terms appear in the form $\log (-\frac{2}{\la})$, see \cite{Zinn-Justin:1981}, \cite{ZJJ:2004}, where $\la$ is the coupling constant. The meaning of the Bogomolny mechanism in superficial terms is in replacement of
$\log (-\frac{2}{\la})$ by $-\frac{1}{2}\log (\frac{\la^2}{4})+ \mbox{const}$} .
It is worth emphasizing that one can see explicitly in (\ref{E-PT-Trans}) the presence
of two structures,
\begin{equation}
\label{structures}
    \xi\ =\ \frac{1}{g} \,e^{-S_0}\ ,\quad \chi\ =\ \log (g^2)\,\frac{1}{g}
    \,e^{-S_0}\ ,
\end{equation}
in addition to the coupling constant $g$ itself, c.f. \cite{ZJJ:2004}, eqs.(8.1)-(8.2). Therefore, the trans-series (\ref{E-PT-Trans}) can be considered as the triple Taylor expansion in $g, \xi, \chi$,
\begin{equation}
\label{newPT}
       E \ =\ \sum E_{k, \ell, p}\, g^{2k} \xi^{\ell} \chi^p\ .
\end{equation}
Note that $\chi$ has a meaning of one-instantion contribution in a leading order: classical action plus determinant. It is worth noting that non-perturbative energy $E_{\rm NPT}$ can be reorganized to the form of perturbation series,
\be
\label{E-NPT-series}
     E_{\rm NPT}\ =\ \sum_{n=0} g^{2n} A_n E_{NPT,n}(g)\ \ ,
\ee
where
\begin{eqnarray}
\label{E-NPT-0}
     E_{\rm NPT,0}\ & = &\  \bigg\{ \frac{1}{g}\,e^{-S_0}\ +\
                       \frac{1}{g^2}\,e^{-2S_0}\,[ A^{(2)}_0\ +\ B^{(2)}_0 \log (g^2) ]  \\
        \ +\ & & \frac{1}{g^3}\,e^{-3S_0}\,[A^{(3)}_0\ +\ B^{(3)}_{0,1} \log (g^2)\ +\ B^{(3)}_{0,2} \log^2 (g^2)]\ +\ \ldots   \non \\
       \ +\ & & \frac{1}{g^p}\,e^{-pS_0}\,\sum_{q=0}^{p-1} A^{(p)}_{0,q} \log^q (g^2)\ +\ \ldots \bigg\} \ ,\non
\end{eqnarray}
with $A_0=-{\sqrt \frac{1}{\pi}}$, it is the leading all-over-instanton contribution to non-perturbative energy, it represents the sum over multi-instanton saddle points in leading approximation, classical action plus determinant (one-loop contribution). The $n$th correction in (\ref{E-NPT-series}) has a similar form,
\begin{eqnarray}
\label{E-NPT-n}
     E_{\rm NPT,n}\ & = &\  \bigg\{ \frac{1}{g}\,e^{-S_0}\ +\
       \frac{1}{g^2}\,e^{-2S_0}\,[ A^{(2)}_n\ +\ B^{(2)}_n \log (g^2) ]  \\
       \ +\ & & \frac{1}{g^3}\,e^{-3S_0}\,[A^{(3)}_n\ +\ B^{(3)}_{n,1} \log (g^2)\ +\
       B^{(3)}_{n,2} \log^2 (g^2)]\ +\ \ldots   \non \\
       \ +\ & & \frac{1}{g^p}\,e^{-pS_0}\,\sum_{q=0}^{p-1} A^{(p)}_{n,q} \log^q (g^2)\ +\ \ldots \bigg\}
       \ . \non
\end{eqnarray}

A natural question to ask is whether does exist trans-series expansion for wavefunction of the type (\ref{newPT}) with $x$-dependent coefficients and if so how to construct it.
In order to proceed let us derive the generalized Bloch equation, c.f. \cite{Escobar-Ruiz:2017},
specific for the potential with two degenerate minima.
The first step is standard, we begin with the Schr\"odinger equation for the wave function and go to one on its logarithmic derivative $y(x)$, which eliminates the overall normalization constant from consideration.
We arrive at the familiar Riccati equation where the boundary condition $y(0)=0$ should be imposed.
However, in order to find the solution which will guarantee the normalizability of the eigenfunction two extra conditions should be imposed: (i) $y$ should be asymptotically antisymmetric, $y(-x)=-y(x)$, in concrete, it behaves asymptotically like $y(x) \sim g\,x|x|$ at large $|x|$ (at $g>0$), and (ii) derivative at origin is equal to the eigenvalue, $y'(0)=E$. The condition (ii) reveals the meaning of quantization of energy
in the non-linear Riccati equation: for given $g$ there exists the single value $E(g)$ for which (i) holds.
The second step is that we have to extract the product of two linear functions of coordinate from the logarithmic derivative assuming the remaining function depends essentially on the classical coordinate $(g\,x)$,
\be
  x\,(1 - g x)\, z(g\,x, g)\ =\ - \frac{\psi'(x)}{\psi(x)}\ =\ y(x) \ .
\label{eqn_log_der}
\ee
It reflects the fact that since the original potential $V(x)$ (\ref{potential})
has two minima at $x=0$ and $x=1/g$ the logarithmic derivative of wavefunction
(the derivative of the phase) has to vanish linearly at $x=0$ and $x=1/g$, respectively.
Now we have to write the equation for function $z$.
Substituting the construction (\ref{eqn_log_der}) to the Schr\"odinger equation

\[
\bigg(\,-\frac{1}{2}\,\frac{d^2}{dx^2} + \frac{1}{g^2}\, \tilde V(g x)\,\bigg)\,\psi(x) \ = \ E\,\psi(x)\ ,
\]
where the Planck constant is placed equal to one, $\hbar=1$, and redefining the coordinate
$u=g\, x$ assuming $g>0$, we arrive at the equation,
\be
  g^2 u (1-u) z'(u)\ +\ g^2 (1-2u) z(u)\ -\ u^2 (1-u)^2 z(u)^2\ =\
  2\, g^2\, E - \tilde V(u)\ ,\ \tilde V(u)=u^2(1-u)^2\ ,
\label{Riccati-Bloch}
\ee
which is called {\it the generalized Bloch equation}. Note, here $z(u)$ has a meaning of
{\it reduced} logarithmic derivative, see (\ref{eqn_log_der}). We will study the equation (\ref{Riccati-Bloch}), imposing the boundary condition $z(0)=E$ and putting also the condition
$z(u) \sim \mp 1$ at $u \rar \pm \infty$.

Now we proceed to solving the equation (\ref{Riccati-Bloch}) at weak coupling regime $g \rar 0$ by expanding consistently both the energy $E$ and $z(u)$ in trans-series (\ref{E-PT-Trans}) and
\be
\label{E-PT-Trans-Z}
     z(u)\ =\ \sum_{n=0} g^{2n} z_{PT,n}(u)\ +
\ee

\[
 g\,e^{-S_0} \, \left(\zeta^{(1)}_0(u) + g^2 \zeta^{(1)}_1(u)  + \ldots \right)\ +\
 e^{-2S_0}\,\left(\zeta^{(2)}_0(u) + g^2 \zeta^{(2)}_1(u) + \ldots \right) + \ldots
\]

\[
 +\ \log g^2 \, \bigg( e^{-2S_0}\, \left(\tilde{\zeta}^{(2)}_0(u) + g^2 \tilde{\zeta}^{(2)}_1(u)  + \ldots \right)\  +\ \frac{1}{g}
 e^{-3S_0}\left(\tilde{\zeta}^{(3)}_0(u) + g^2 \tilde{\zeta}^{(3)}_1(u)  + \ldots\right) + \ldots\bigg)\ +
\]

\[
 +\ \frac{1}{g} \log^2 g^2 \, \bigg( e^{-3S_0} \left(\hat{\zeta}^{(3)}_0 + g^2 \hat{\zeta}^{(3)}_1 + \ldots \right)  + \frac{1}{g^2} e^{-4S_0} \left(\hat{\zeta}^{(4)}_0 + g^2 \hat{\zeta}^{(4)}_1 + \ldots \right) + \ldots\bigg)\ + \ldots\ ,
\]
respectively. We will explore in details the following issues: (i) the perturbation theory in powers of $g$, (ii) the one-instanton contribution $\sim e^{-S_0}$ and (iii) two-instanton contributions $\sim e^{-2S_0}$,
and (iv) the sum of leading multi-instanton contributions.

\section{Weak coupling regime: perturbation series vs semiclassical expansion}

Looking at the generalized Bloch equation (\ref{Riccati-Bloch}) one can immediately realize a striking fact that the perturbation theory expansion
\be
\label{EptZpt}
   E_{PT}(g)\ =\ \sum_{n=0} g^{2n} E_{PT,n}\ ,\ z_{PT}(u)\ =\ \sum_{n=0} g^{2n} z_{PT,n}(u)\ ,
\ee
can be constructed self-consistently, without involving non-perturbative, exponentially small terms, c.f. \cite{Escobar-Ruiz:2017}, Section III.C.2\,. Owing to this property we can {\it separate} perturbative and non-perturbative contributions in $z$!
Since now on we will drop the notation ``PT'' in $z$ but will keep it for energy $E$.

In the zeroth order in $g$, $O(g^0)$ in (\ref{Riccati-Bloch}), in which all terms proportional to the coupling are ignored, the equation to solve is very simple
\be
\label{Ez0}
   -\, u^2 (1-u)^2 \,z_0(u)^2\ =\ - u^2(1-u)^2\ ,
\ee
leading to
\be
\label{z0}
     z_0(u)\ =\ \pm 1 \ ,
\ee
here the sign is chosen by requiring the normalizability of the unperturbed wave function $\Psi_0$.
It will be taken the sign plus for $u<0$: $z_0=-1$  and the sign minus for $u>0$: $z_0=1$. Hence, the solution is discontinuous at $u=0$. This is the indication that we can not go to domain of small $|u|$: the radius of convergence $u_0 > 0$ of the expansion (\ref{EptZpt}) for $z_{PT}$ is finite: $|u| > u_0$, see below.

This result $(u(1-u)\, z_0)$ is, in fact, the classical momentum at zero energy, and therefore,
when we return to the wave function, the zeroth order term gives the well known semiclassical action.
So, zero approximation admits a simple interpretation as the exponent = classical action in semiclassical wavefunction $\psi \sim exp(-\int^x p(x') dx')$ but at zero energy.

Moving to the next term of the expansion, one finds the following equation $O(g^2)$ for it
\be
\label{Ez1}
u (1-u) z_0'(u)\ +\ (1-2u) z_0(u)\ -\ 2 \, u^2 (1-u)^2 z_0(u) z_1(u)\ =\ 2\,E_{PT,0}\ .
\ee
Note here, that the equation involves the known function $z_0$ and unknown $z_1$, both of them appear linearly.
The similar feature takes place in all orders(!): finding $z_n$ does not involve solving a differential equation rather than a linear algebraic one.

Important feature of the procedure is that the perturbative energy $E_{PT}$ needs to be used in (\ref{Riccati-Bloch}) instead of $E$, in the form of perturbative expansion in powers of $g^2$.
These coefficients $E_{PT,n}$ should be found separately, by some other method,
not via the perturbation theory in generalized Bloch equation.
For example, non-linearization procedure can be used for it \cite{Turbiner:1984}.
Since the zeroth order potential is the harmonic oscillator one, so $E_{PT,0}=1/2$.
Hence, the first correction, which emerges from (\ref{Ez1}), is given by
\be
\label{z1}
   z_1(u)\ =\ \frac{(1-2u) z_0 -1 }{2 u^2 (1-u)^2 z_0}\ ,
\ee
which is rational function in $u$. At large $u > 0$ the correction tends to zero, $z_1 \rar -u^{-3}$ in agreement with boundary conditions at large $|u|$. Otherwise, it grows up to infinity with decreasing $|u|$ towards zero or one. It implies that we can not go to domain of small $|u|$ and should remain at large $|u|$, which is typical for semiclassical approximation. In \cite{Escobar-Ruiz:2017} it was shown explicitly that this correction is related to the determinant in flucton loop expansion. In similar way one can find $z_2(u)$ using the first perturbation correction $E_{PT,1}$ and known $z_{0,1}$ by solving the equation
\be
\label{Ez2}
u (1-u) z_1'(u)\ +\ (1-2u) z_1(u)\ -\  u^2 (1-u)^2 (z_1^2 + 2 \,z_0(u) z_2(u))\ =\ 2\,E_{PT,1}\ .
\ee
As the result
\be
\label{z2}
   z_2(u)\ =\ \frac{u (1-u) z_1'(u)+ (1-2u) z_1 - 2\,E_{PT,1} }{2 u^2 (1-u)^2 z_0} - \frac{z_1^2}{2 z_0}\ ,
\ee
is the rational function in $u$. At $|u| \rar \infty$, $z_2 \sim -\frac{3}{2u^6}$, overall, it is of the order $O(g^4)$. This correction is related with two-loop contribution in flucton loop expansion \cite{Escobar-Ruiz:2017}.

In general, in the same way one can write the equation for $z_n(u)$
\be
\label{Ezn}
u (1-u) z_{n-1}'(u)\ +\ (1-2u) z_{n-1}(u)\ -\  u^2 (1-u)^2 [Q_{n} + 2 \,z_0(u) z_n(u)]\ =\ 2\,E_{PT,{n-1}}
\ ,
\ee
where
\[
  Q_{n}\ =\ \sum_{i=1}^{n-1} \ z_{i} z_{n-i} \ .
\]
Finally, the solution gets the form
\be
\label{zn}
   z_n(u)\ =\ \frac{u (1-u) z_{n-1}'(u)+ (1-2u) z_{n-1} - 2\,E_{PT,n-1} }{2 u^2 (1-u)^2 z_0} - \frac{Q_n}{2 z_0}\ .
\ee
In general, it is the rational function in u,
\be
\label{zn-rat}
   z_n(u)\ =\ \frac{p_n(u)}{2 u^{2n} (1-u)^{2n} z_0^{n}} \ =\ \frac{p_n(u)}{2 \tilde V^n} \ ,
\ee
where $p_n$ is the $n$th degree polynomial with rational coefficients and $\tilde V$ is the potential defined (\ref{Riccati-Bloch}). Thus, $z_n(u)$ is given by the sum of $n$-loop Feynman diagrams weighted with appropriate symmetry factors in flucton calculus.

\section{Weak coupling regime: trans-series expansion, exponentially-small terms}

\subsection{One-instanton contribution}

Analysing the generalized Bloch equation (\ref{Riccati-Bloch}) one can immediately realize a striking fact that the one-instanton contribution
\be
\label{E-1I-Z-1I}
   E_{1I}(g)\ =\ A_0 e^{-S_0} \sum_{n=0} g^{2n-1} E_{1I,n}\ ,\ z_{1I}(u)\ =\ A_0 e^{-S_0}\sum_{n=0} g^{2n+1} \zeta^{(1)}_n(u)\ ,
\ee
see (\ref{E-PT-Trans}), 2nd line, here $A_0=-{\sqrt \frac{1}{\pi}}$ is normalization factor given by the instanton determinant at $g=1$ and $E_{1I,n}$ define energy corrections to one-instanton; systematically, they are rational numbers $E_{1I,0}=1, E_{1I,1}=-\frac{71}{12}, E_{1I,2}=-\frac{6299}{288} \ldots $ \cite{ZJJ:2004}, Section 8, Eq.(8.13a); and see (\ref{E-PT-Trans-Z}), 2nd line can be constructed without involving exponentially small terms of higher orders $e^{- p S_0}, p \geq 2$. Note that $E_{1I,n}$ at $n=1,2$ were calculated alternatively in instanton calculus using 2- and 3-loop Feynman integrals \cite{E.Shuryak,EST-I}, respectively.

Now we proceed to calculation of exponentially-small terms in $g$ in expansion (\ref{E-PT-Trans-Z}), (\ref{E-1I-Z-1I}). As the first step let us collect all terms of the order $O(g\,e^{-S_0})$ in eq.(\ref{Riccati-Bloch}), which is of the lowest order in $g$ in front of the exponentially-small term $e^{-S_0}$,
\be
\label{Ezeta1}
-\ 2 \, u^2 (1-u)^2 z_0(u)\, \zeta_{0}^{(1)}(u)\ =\ -2\,E_{1I,0}\ ,
\ee
c.f.(\ref{Ez0}), where $E_{1I,0}=1$, see e.g. \cite{Zinn-Justin:1981} and $z_0$ is given by (\ref{z0}). Its solution has the form,
\be
\label{zeta_1}
  \zeta_{0}^{(1)}\ =\ \frac{1}{u^2(1-u)^2\,z_0}\ =\ \frac{1}{\tilde V z_0}\ ,
\ee
for $|u| > 1$, here the potential $\tilde V$ is defined at (\ref{Riccati-Bloch}).
Asymptotically,
\be
\label{zeta-1}
  \zeta_{0}^{(1)}\ \rar \   \frac{1}{z_0 u^4}\ ,\quad u \rar \pm \infty \ ,
\ee
hence, the boundary condition at $u=\pm \infty$ is satisfied.
As the next step let us collect all terms of the order $O(g^3\,e^{-S_0})$ in eq.(\ref{Riccati-Bloch}), which is of the next-to-lowest order in $g$ in front of the exponentially-small term $e^{-S_0}$,
\[
u (1-u)\, (\pa_u{\zeta_{0}^{(1)}}(u))\ +\ (1-2u)\, \zeta_{0}^{(1)}(u)\ -
\ 2 \, u^2 (1-u)^2 z_0\, \zeta_{1}^{(1)}(u)\ =\
\]
\be
\label{Ezeta_2}
  -2\,E_{1I,1} + \ 2 \, u^2 (1-u)^2 z_1\, \zeta_{0}^{(1)}(u) \ ,
\ee
where $E_{1I,1}=-71/12$, see e.g. \cite{Zinn-Justin:1981} and also \cite{E.Shuryak} and $z_1$ is given by (\ref{z1}). Its solution has the form,
\be
\label{zeta_2}
  \zeta_{1}^{(1)}\ =\ \frac{1}{z_0}\,
  \bigg( -\frac{71}{12 u^2(1-u)^2}\ -\ z_1\,\zeta_{0}^{(1)} + \frac{\pa_u{\zeta_{0}^{(1)}}}{2 u (1-u) } + \frac{(1-2u)\, \zeta_{0}^{(1)}}{2 u^2(1-u)^2}
  \bigg)
  \ .
\ee

In general, collecting terms of the order $O(g^{2n+1}\,e^{-S_0})$ in eq.(\ref{Riccati-Bloch}), we arrive at the equation
\[
u (1-u)\, (\pa_u{\zeta_{n-1}^{(1)}}(u))\ +\ (1-2u)\, \zeta_{n-1}^{(1)}(u)\ -
\ 2 \, u^2 (1-u)^2 z_0\, \zeta_{n}^{(1)}(u)\ =\
\]
\be
\label{Ezeta_n}
  -2\,E_{1I,n} + \ 2 \, u^2 (1-u)^2 Q^{(1)}_n \ ,
\ee
where
\[
   Q^{(1)}_n\ =\ \sum_{i=1}^{n} z_i(u)\,\zeta_{n-i}^{(1)}(u)\ .
\]
It is easily solved and the explicit form of the $n$th correction reads,
\be
\label{zeta_n}
  \zeta_{n}^{(1)}\ =\ \frac{1}{z_0}\,
  \bigg( -\frac{E_{1I,n}}{ u^2(1-u)^2}\ -\ Q^{(1)}_n\ +\ \frac{\pa_u{\zeta_{n-1}^{(1)}}}{2 u (1-u) } + \frac{(1-2u)\, \zeta_{n-1}^{(1)}}{2 u^2(1-u)^2}
  \bigg)
  \ .
\ee
Finally, the $n$th correction has the form of a rational function with integer coefficients similar to
(\ref{zn-rat}).

Concluding one can see that in order to construct $z_{1I}(u)$ we have to know perturbative contribution $z_{PT}(u)$ only. It is a type of nested construction.

\subsection{Two-instanton contribution}

From the generalized Bloch equation (\ref{Riccati-Bloch}) one can immediately realize that the two-instanton contribution
\be
\label{E-2I-Z-2I}
   E_{2I}(g)\ =\ A^{(2)}_0 e^{-2S_0} \sum_{n=0} g^{2n-2} E_{2I,n}\ ,\ z_{2I}(u)\ =\ A^{(2)}_0 e^{-2S_0}\sum_{n=0} g^{2n} \zeta^{(2)}_n(u)\ ,
\ee
see (\ref{E-PT-Trans}), 2nd line, and, see (\ref{E-PT-Trans-Z}), 2nd line, can be constructed without involving exponentially small terms of higher orders $e^{- p S_0}, p > 2$ or logarithmic contributions
$\log^q (g^2) e^{- p S_0}, q \geq 1, p \geq 2$.

Here $A^{(2)}_0={\frac{1}{\pi}}$ is normalization factor given seemingly by the two-instanton determinant at $g=1$ and $E_{2I,n}$ define energy corrections to two-instanton,
systematically, they are written in the form of linear function in Euler constant $\gamma$ with rational coefficients:
\[
E_{2I,0}=\gamma\ ,\ E_{2I,1}= -\frac{23}{2} - \frac{53}{6}\gamma\ ,\ E_{2I,2}=\frac{13}{12} - \frac{1277}{72}\gamma \ldots\ ,
\]
see \cite{ZJJ:2004}, Section 8, Eq.(8.14a). We are not familiar with any attempt to calculate these coefficients in instanton calculus.

Collecting the terms of the order $O(g^0 \, e^{-2S_0})$ in eq.(\ref{Riccati-Bloch}), which is the lowest order in $g$ in front of the exponentially-small term $e^{-2S_0}$, we arrive at
\be
\label{Ezeta-2}
-\ 2 \, u^2 (1-u)^2 z_0(u)\, \zeta_{0}^{(2)}(u)\ =\ -2\,E_{2I,0}\ ,
\ee
c.f.(\ref{Ez0}), where $E_{2I,0}=1$, see e.g. \cite{Zinn-Justin:1981} and $z_0$ is given by (\ref{z0}). Its solution has the form,
\be
\label{zeta-2}
  \zeta_{0}^{(2)}\ =\ \frac{1}{u^2(1-u)^2\,z_0}\ =\ \frac{1}{\tilde V z_0}\ ,
\ee
for $|u| > 1$, here the potential $\tilde V$ is defined at (\ref{Riccati-Bloch}). It coincides with $\zeta_{0}^{(1)}$ (\ref{zeta_1}).

As the next step let us collect all terms of the order $O(g^2\,e^{-2S_0})$ in eq.(\ref{Riccati-Bloch}), which is of the next-to-lowest order in $g$ in front of the exponentially-small term $e^{-2S_0}$,
\[
u (1-u)\, (\pa_u{\zeta_{0}^{(2)}}(u))\ +\ (1-2u)\, \zeta_{0}^{(2)}(u)\ -
\ 2 \, u^2 (1-u)^2 z_0\, \zeta_{1}^{(2)}(u)\ =\
\]
\be
\label{Ezeta-2-1}
  -2\,E_{2I,1} + \  u^2 (1-u)^2 \left(2\,z_1\, \zeta_{0}^{(2)} + (\zeta_{0}^{(1)})^2\right)\ ,
\ee
where $z_1$ is given by (\ref{z1}) and $\zeta_{0}^{(1)}$ is from (\ref{zeta_1}). Its solution has the form,
\be
\label{zeta-2-2}
  \zeta_{1}^{(2)}\ =\ \frac{1}{z_0}\,
  \bigg( \frac{E_{2I,1}}{ u^2(1-u)^2}\ -\ \frac{1}{2}\left(2\,z_1\, \zeta_{0}^{(2)} + (\zeta_{0}^{(1)})^2\right) + \frac{\pa_u{\zeta_{0}^{(2)}}}{2 u (1-u) } + \frac{(1-2u)\, \zeta_{0}^{(2)}}{2 u^2(1-u)^2}
  \bigg)
  \ .
\ee
It is easy to find the $n$th correction
\be
\label{zeta-2-n}
  \zeta_{n}^{(2)}\ =\ \frac{1}{z_0}\,
  \bigg( \frac{E_{2I,n}}{ u^2(1-u)^2}\ -\ \frac{Q^{(2)}_n}{2} + \frac{\pa_u{\zeta_{n-1}^{(2)}}}{2 u (1-u) } + \frac{(1-2u)\, \zeta_{n-1}^{(2)}}{2 u^2(1-u)^2}
  \bigg)
  \ .
\ee
where
\[
   Q^{(2)}_n\ =\ 2\sum_{i=1}^{n} z_i(u)\,\zeta_{n-i}^{(2)}(u)\ +\ \sum_{i=0}^{n-1} \zeta_{i}^{(1)}(u)\,\zeta_{n-i}^{(1)}(u)\ .
\]

It is evident that in order to construct two-instanton contribution $z_{2I}(u)$ we have to know perturbative contribution $z_{PT}(u)$ and one-instanton contribution $z_{1I}(u)$ only. As a result
the correction $\zeta_{n}^{(2)}$ is a rational function

Needless to demonstrate that in order to determine the $k$-instanton contribution,
\be
\label{E-kI-Z-kI}
   E_{kI}(g)\ =\ A^{(k)}_0 e^{-2S_0} \sum_{n=0} g^{2n-k} E_{2I,n}\ ,\ z_{kI}(u)\ =\ A^{(k)}_0 e^{-2S_0}\sum_{n=0} g^{2n-k+2} \zeta^{(2)}_n(u)\ ,
\ee
we have to know perturbative contribution $z_{PT}(u)$ and all one-, two-, $(k-1)$-instanton contributions $z_{(k-1)I}(u)$. It is a type of nested construction, it does not involve logarithmic contributions .

\subsection{Two-instanton log contribution}

From the generalized Bloch equation (\ref{Riccati-Bloch}) one can immediately realize that the two-instanton contribution
\be
\label{E-2I-Z-2I-log}
   E_{2I-log}(g)\ =\ A^{(2l)}_0\, \log (g^2)\, e^{-2S_0}\, \sum_{n=0} g^{2n-2} E_{2Il,n}\ ,\ z_{2Il}(u)\ =\ A^{(2l)}_0\,\log (g^2)\, e^{-2S_0}\sum_{n=0} g^{2n} \zeta^{(2l)}_n(u)\ ,
\ee
see (\ref{E-PT-Trans}), 2nd line, and, see (\ref{E-PT-Trans-Z}), 2nd line, can be constructed without involving exponentially small terms of higher orders $e^{- p S_0}, p > 2$ or logarithmic contributions
$\log^q (g^2) e^{- p S_0}, q \geq 1, p > 2$.

Here $A^{(2l)}_0={\frac{1}{\pi}}$ is normalization factor given seemingly by the two-instanton determinant at $g=1$ and $E_{2Il,n}$ define energy corrections to two-instanton logarithmic contribution,
systematically, they are given by rational coefficients:
\[
E_{2Il,0}\ =\ 1\ ,\ E_{2Il,1}\ =\ -\frac{53}{6}\ ,\ E_{2Il,2}\ =\ \frac{1277}{72}\ ,\  \ldots\ ,
\]
see \cite{ZJJ:2004}, Section 8, Eq.(8.14a). We are not familiar with any attempt to calculate these coefficients in instanton calculus.

Collecting the terms of the order $O(\log (g^2) \, e^{-2S_0})$ in eq.(\ref{Riccati-Bloch}), which is the lowest order in $g$ in front of the exponentially-small term $\log (g^2) e^{-2S_0}$, we arrive at
\be
\label{Ezeta2l}
-\ 2 \, u^2 (1-u)^2 z_0(u)\, \zeta_{0}^{(2l)}(u)\ =\ -2\,E_{2Il,0}\ ,
\ee
c.f.(\ref{Ez0}), where $E_{2I,0}=1$, see e.g. \cite{Zinn-Justin:1981} and $z_0$ is given by (\ref{z0}). Its solution has the form,
\be
\label{zeta2l}
  \zeta_{0}^{(2l)}\ =\ \frac{1}{u^2(1-u)^2\,z_0}\ =\ \frac{1}{\tilde V z_0}\ ,
\ee
for $|u| > 1$, here the potential $\tilde V$ is defined at (\ref{Riccati-Bloch}).
It coincides with $\zeta_{0}^{(1)}$ (\ref{zeta_1}) and with $\zeta_{0}^{(2)}$ (\ref{zeta-2}).

It is easy to find the $n$th correction
\be
\label{zeta2l-2}
  \zeta_{n}^{(2l)}\ =\ \frac{1}{z_0}\,
  \bigg( \frac{E_{2I,n}}{ u^2(1-u)^2}\ -\ \frac{Q^{(2l)}_n}{2} + \frac{\pa_u{\zeta_{n-1}^{(2l)}}}{2 u (1-u) } + \frac{(1-2u)\, \zeta_{n-1}^{(2l)}}{2 u^2(1-u)^2}
  \bigg)
  \ .
\ee
where
\[
   Q^{(2l)}_n\ =\ 2\sum_{i=1}^{n} z_i(u)\,\zeta_{n-i}^{(2)}(u)\ .
\]
One can see that in order to construct $\zeta_{2Il}(u)$ we have to know perturbative contribution $z_{PT}(u)$ only. Thus, it is a type of nested construction. As a result
the correction $\zeta_{n}^{(2l)}$ is a rational function

\subsection{Leading semiclassical multi-instanton-inspired correction}

The sum of the exponentially small contributions to the ground state energy in the leading order, when the perturbation theory around multi-instanton is neglected, can be written in the form
\[
 A_0\, E_{NPT,0}\ =
\]
\be
\label{Ezeta-lead}
 \ A_0 \bigg(
  \sum_{p=1} B_0^{(l^0,p)}  g^{-p} e^{-pS_0} + \log (g^2)\sum_{p=2} B_0^{(l,p)}  g^{-p} e^{-pS_0} +
   \log^2 (g^2)\sum_{p=3} B_0^{(l^2,p)}  g^{-p} e^{-pS_0} + \ldots \bigg)\ .
\ee
c.f. (\ref{E-NPT-0}).
We assume and then check correctness afterwards that the sum of the exponentially small contributions in $g$ to the reduced phase $z$ in the leading order, when the perturbation theory around multi-instanton is neglected, has the form,
\[
 A_0\, \zeta_{NPT,0}(u)\ =\
\]
\be
\label{zeta-lead}
  A_0\, \bigg( \sum_{p=1} B_0^{(l^0,p)}  g^{-p+2} e^{-pS_0}\,\zeta^{(l^0,p)}_0(u)\ +
\ee
\[
   \log (g^2)\sum_{p=2} B_0^{(l,p)}  g^{-p+2} e^{-pS_0}\,\zeta^{(l,p)}_0(u)\ +\
   \log^2 (g^2)\sum_{p=3} B_0^{(l^2,p)}  g^{-p+2} e^{-pS_0}\,\zeta^{(l^2,p)}_0(u)\ +\ \ldots \bigg)\ ,
\]
where $l^q$ in superscript of $B_0^{(l^2,p)}$ means presence of the $\log^q$ in front of sum.

Now let us take the generalized Bloch equation (\ref{Riccati-Bloch}), substitute in there the energy in the form (\ref{E-PT-Trans}) and the reduced logarithmic derivative $z(u,g)$ in the form (\ref{E-PT-Trans-Z}), and collect carefully, one by one, the expressions in $g$ and $e^{-S_0}$ which occur in (\ref{zeta-lead}). Finally, it turns out that the coefficient in front of the defining expression has the form
\be
\label{Ezeta-lead1}
-\ 2 \, u^2 (1-u)^2 z_0(u)\, \zeta_{0}(u)\ + \ 2\,B_{0}\ =\ 0\ ,
\ee
(where upper indices in $\zeta_0$ and $B_0$ are dropped for convenience) independently on upper indices,
c.f.(\ref{Ez0}) as well as (\ref{Ezeta1}), (\ref{Ezeta-2}), (\ref{Ezeta2l}), here $z_0$ is given by (\ref{z0}). Its solution has the form,
\be
\label{zeta-lead0}
  \zeta_{0}(u)\ =\ \frac{B_0}{u^2(1-u)^2\,z_0}\ =\ \frac{B_0}{\tilde V z_0}\ ,
\ee
for $|u| > 1$, here the potential $\tilde V$ is defined at (\ref{Riccati-Bloch}). Substituting (\ref{zeta-lead0}) into (\ref{zeta-lead}) we arrive at unexpectedly compact expression,
\be
\label{zeta-lead-final}
  \zeta_{NPT,0}(u)\ =\ \frac{E_{NPT,0}}{u^2(1-u)^2\,z_0}\ =\ \frac{E_{NPT,0}}{\tilde V(u) z_0}\ .
\ee
It corresponds to logarithmic derivative
\[
     y_{NPT,0}\ =\ x(1-gx)\,\zeta_{NPT,0}(gx)\ =\ \frac{1}{g^2}\ \frac{E_{NPT,0}}{\sqrt {V(x))} z_0}
\]
and the non-perturbative phase at large $x \gg 0$ is equal to
\be
\label{NPT-0}
     \phi_{NPT,0}(x) \ =\ \frac{E_{NPT,0}}{g^2 z_0}\ \int \frac{1}{\sqrt {V(x)}} dx\ =\
     -\frac{E_{NPT,0}}{g^2} \log (1 - \frac{1}{gx })\ \approx \frac{E_{NPT,0}}{g^2}\ \frac{1}{gx}\ .
\ee
Hence, the non-perturbative phase is subdominant in comparison to the classical action in semiclassical phase, which is leading (dominant) contribution,
\be
\label{PT-0}
    \phi_{PT,0}(x)(x)  \ =\ g \frac{x^3}{3} - \frac{x^2}{2}\ ,\ x \gg 0 \ ,
\ee
also the first perturbative correction, which is next-to-leading contribution
\be
\label{PT-1}
   \phi_{PT,1}(x)(x)  \ =\ \log gx   \ .
\ee
see (\ref{z1}). However, the second perturbative correction (\ref{z2}) which leads to next-to-next-to-leading contribution,
\be
\label{PT-2}
   \phi_{PT,2}(x)(x)  \ =\ -\frac{9}{2g}\ \frac{1}{x^3}   \ ,
\ee
is subdominant to leading non-perturbative correction (\ref{NPT-0}). Hence, non-perturbative correction (\ref{NPT-0}) being of order $O(1/x)$ provides asymptotic behavior intermediate to the first and second perturbative corrections
being of ``alien" nature for semi-classical perturbation theory. It can be used to calculate the leading non-perturbative instantonic contribution to the energy gap $E_{NPT,0}$ as a coefficient in front of $1/x$ term in asymptotic expansion of phase.

Following the philosophy of construction of approximate wave function for double-well potential
\cite{Turbiner:2005},\cite{Turbiner:2010} neither leading non-perturbative correction $\phi_{NPT,0}(x)$, nor the perturbative correction $\phi_{PT2,0}(x)$ are of importance.

\section{Connection to path integrals}

Now when the perturbative and non-perturbative corrections to the phase of wave function in semi-classical perturbation theory are found, we would like to return to the original issue indicated in Introduction: the contributions of the
flucton and flucton-plus-instanton classical path contributions should naturally appear additively
in the path integral for the density matrix $\rho(x_0)=\psi(x_0)^2=\exp{(-2\phi(x_0))}$.
In order to do it the representation (\ref{psiPT+psiNPT}) used to construct trans-series expansion should be rewritten as the product of two factors
\[
   \Psi\ =\ e^{-\phi_{\rm PT}\ -\ \phi_{\rm NPT}} \ =\ e^{-\phi_{\rm PT}}\ e^{-\ \phi_{\rm NPT}}\ .
\]
As we already know \cite{Escobar-Ruiz:2016}, \cite{Escobar-Ruiz:2017} the Taylor expansion
\[
 \phi_{\rm PT}(u)\ =\ \sum_{n=0} g^{2n} \phi_{PT,n}(u)\ ,
\]
corresponds to the loop expansion in flucton calculus: $\phi_{PT,0}$ is classical flucton action,
one-loop contribution $g^2\phi_{PT,1}=\log D$ is logarithm of determinant, $\phi_{PT,2}$ is two-loop
contribution and, in general, $\phi_{PT,n}$ is $n$-loop contribution. It allows us to rewrite
the perturbative part of the flucton density
$(\Psi_{\rm PT})^2$ as the saddle-point expansion,
\be
\label{PT-saddle}
  e^{-2\phi_{\rm PT}}\ =\ e^{-2 \phi_{PT,0}}\,F_0\ \equiv \ \frac{1}{D^2}\,e^{-2 \phi_{PT,0}}
  (1 - 2 g^4 \phi_{PT,2} + \ldots)
\ee
The second factor $e^{-\ 2\phi_{\rm NPT}}$ can be expanded in the Taylor series in powers of non-perturbative phase $\phi_{\rm NPT}$. It corresponds to the expansion in powers of the exponential in one-instanton classical action,
\be
\label{NPT-saddle}
  e^{-2\phi_{\rm NPT}}\ =\ 1\ +\ e^{-S_0} F_1 (x, g) \ +\ e^{-2 S_0} F_2 (x, g) \ +\ \ldots\ \ ,
\ee
where for functions $F_{1,2,\ldots}$ the first terms in the expansion in powers $g$ can be found explicitly.
In particular,
\[
  F_1 \ =\ \int \frac{dx}{gx(1-gx) z_0}\bigg(1 - \frac{g^2}{2}\bigg(\frac{83}{6}+\frac{1+2gx}{(gx)^2 (1-gx)^2 z_0}\bigg)+ \dots \bigg)
\]
Thus, the expansion (\ref{NPT-saddle}) appears as the expansion in powers $e^{-S_0}$. Combining (\ref{PT-saddle}) and (\ref{NPT-saddle}) we arrive at the expansion for density in the form a superposition of saddle-point contributions (and expansion around each of them multiplied by the product of determinants),
\be
\label{Saddle-point-expansion}
    e^{-2 \phi}\ =\ e^{-2 \phi_{PT,0}}\,F_0\ +\ e^{-2 \phi_{PT,0} - S_0}\,F_0\,F_1 \ +\ \ldots\ +\
    e^{-2 \phi_{PT,0} - n S_0}\,F_0\,\tilde F_n\ +\ \ldots \ ,
\ee
where $\tilde F_n$ is a polynomial in $F$'s.
The first term corresponds to flucton classical trajectory with classical action $(2 \phi_{PT,0})$ while $\lim_{g\rar 0} \frac{1}{g^2}F_0(u,g)$ represents the determinant 
(quadratic fluctuations), the second one is the flucton+instanton trajectory contribution 
with classical action $(2 \phi_{PT,0}+S_0)$ while $\lim_{g\rar 0} \frac{1}{g^4}F_0(u,g) F_1(u,g)$ represents determinant (quadratic fluctuations) around this trajectory, and the $(n+1)$th term should correspond to flucton+$n$-instanton contribution with classical action $(2 \phi_{PT,0}+n\,S_0)$ while $F_0(u,g=0) \tilde F_n(u,g=0)$ represents determinant (quadratic fluctuations) around this trajectory etc.
The main point is that different classical paths lead to $additive$ contributions 
to the path integral, and thus to the density matrix. It is evident that the expansion (\ref{Saddle-point-expansion}) is different for large positive and negative $x$. 
They correspond to the expansion of the first (second) term in (\ref{e-DWP}), respectively. 
The symmetry is restored when the new variable is introduced $\tilde x=x - \frac{1}{2g}$: 
the expansions become the same.

Furthermore, more close focus on the obtained result reveals one more
interesting phenomenon:
the interaction between classical objects, which lead to logarithmic
terms in the trans-series.
Indeed, let us look again at the lowest order perturbative and non-perturbative results
we already obtained above. The equation reads
\be
u^2(1-u)^2 \big( z_0 + z_{np}\big)={A_0\over g} exp(-S_i)+...
\ee
Using the definitions of $z$ and $u$, it means that
\be -{\psi'_x(x) \over \psi(x)} = x(1-g x)z(u)=x (1-gx) +{1\over x(1-gx)}{A_0\over g^3}exp(-S_i)+... \ee
Integrating over coordinate to recover the wave function one  finds
\be \psi(x)\sim exp\big( - \int_{x_{min}}^x dx' \big[x' (1-gx') +{1\over x'(1-gx')}{A_0\over g^3}exp(-S_i) \big]\big) \approx \ee
$$ exp(-S_f(x_0)) \big[1+ log\big(
{ x(1-g x_{min}) \over x_{min}(1- g x )}
\big){A_0\over g^3} exp(-S_i) +... \big]
$$
where $x > x_{min} $, which is some normalization point. While the flucton and
instanton actions in the second term appear in exponent as a sum,
the pre-exponent has nontrivial logarithmic dependence on $x$ and $g$.
So, the classical flucton and instanton actions are additive, but the determinants do not simply factorize, but   indicate instead appearance of new  series with logs.

In the language of paths this dependence comes from the fact that there is no just one single
``f+i" trajectory, but a whole family of such paths, parameterized by the time $\Delta \tau$ between
their centers. Integration over all paths of the family, over  $\Delta \tau$, is the source
of the discussed interaction. Unfortunately, it is not so simple to calculate
explicitly its effect in the path integral formalism.
But we do not have to do so: we have already found the total contribution of ``f+i"
family of paths.

We therefore reached the main goal of the paper: we indeed see additive contributions
to the density matrix  of the two paths sketched in Fig.1, the flucton one and the flucton-plus-instanton one. One can find in the exponent the simple sum of both actions: this indicates
that generically the flucton and instanton parts of the path are far away and classically do not interact.
However, the pre-exponent does depend on $x$: so at one-loop level such interaction between them does exist. Note that the integral produces logarithms,
of similar origin as inter-instanton logarithms in the transseries for the energy.
The relative normalization of the two (or more, with multi-instantons) contributions
is therefore established.

Finally, we remind the reader that our ultimate goal is to use semiclassical theory of fluctons and instantons in the QFT settings, in which the same issue of relative normalization
is present, but there is no handy generalized Bloch equation available.

\begin{acknowledgments}
E.S. thanks G Dunne to indicating the work \cite{Schulman}.
The work of E.S. is supported in part by the U.S. Department of Energy, Office of Science under Contract
No. {\bf DE-FG-88ER40388}.
A.V.T. gratefully acknowledges support from the Simons Center for Geometry and Physics,
Stony Brook University at which the research for this paper was initiated and eventually
completed.


\end{acknowledgments}

\pagebreak

\end{document}